\documentclass{SCIS2025}
\usepackage{arydshln}
\usepackage{amssymb}
\usepackage{amsmath}
\usepackage{subcaption}
\usepackage{xcolor}

\begin{document}

\ArticleType{RESEARCH PAPER}
\Year{2025}
\Month{}
\Vol{}
\No{}
\DOI{}
\ArtNo{}
\ReceiveDate{}
\ReviseDate{}
\AcceptDate{}
\OnlineDate{}
\AuthorMark{}
\AuthorCitation{}

\title{Topology-driven quantum architecture search framework}{}

\author[1,2,3]{Junjian Su}{}
\author[1]{Jiacheng Fan}{}
\author[1]{Shengyao Wu}{}
\author[1]{Guanghui Li}{}
\author[1]{Sujuan Qin}{}
\author[1]{Fei Gao}{{gaof@bupt.edu.cn}}

\address[1]{State Key Laboratory of Networking and Switching Technology, Beijing University of Posts and Telecommunications,\\ Beijing 100876, china}
\address[2]{National Engineering Research Center of Disaster Backup and Recovery, Beijing University of Posts and Telecommunications,\\ Beijing 100876, china}
\address[3]{School of Cyberspace Security, Beijing University of Posts and Telecommunications, Beijing 100876, china}

\abstract{
The limitations of Noisy Intermediate-Scale Quantum (NISQ) devices have motivated the development of Variational Quantum Algorithms (VQAs), which are designed to potentially achieve quantum advantage for specific tasks. Quantum Architecture Search (QAS) algorithms play a critical role in automating the design of high-performance Parameterized Quantum Circuits (PQCs) for VQAs. However, existing QAS approaches struggle with large search spaces, leading to substantial computational overhead when optimizing large-scale quantum circuits.
Extensive empirical analysis reveals that circuit topology has a greater impact on quantum circuit performance than gate types. Based on this insight, we propose the Topology-Driven Quantum Architecture Search (TD-QAS) framework, which first identifies optimal circuit topologies and then fine-tunes the gate types.
In the fine-tuning phase, the QAS inherits parameters from the topology search phase, eliminating the need for training from scratch. By decoupling the large search space into separate topology and gate-type components, TD-QAS avoids exploring gate configurations within low-performance topologies, thereby significantly reducing computational complexity. 
Numerical simulations across various tasks, under both noiseless and noisy conditions, validate the effectiveness of the TD-QAS framework. 
This framework advances standard QAS algorithms by enabling the identification of high-performance quantum circuits while minimizing computational demands.
These findings indicate that TD-QAS deepens our understanding of VQAs and offers broad potential for the development of future QAS algorithms.
}


\maketitle

\section{Introduction}
\label{sec_Introduction}
Quantum computing exhibits promising potential in addressing complex problems, such as combinatorial optimization \cite{1, 2}, factoring \cite{3}, linear system solving \cite{4}, chemical simulations \cite{5}, and machine learning \cite{6, 7, 8, 9, 10, 11, 12}. However, Noisy Intermediate-Scale Quantum (NISQ) devices \cite{13}, which represent the forefront of near-term quantum technology, face substantial challenges, such as noise, limited qubit count, and restricted qubit connectivity. These limitations impede the widespread application of quantum computing \cite{14}. Therefore, Variational Quantum Algorithms (VQAs) have been developed to operate within the constraints of NISQ devices \cite{15, 16, 17, 18}, offering the potential to achieve quantum advantage for specific tasks.

The effectiveness of Variational Quantum Algorithms (VQAs) highly depends on the design of Parameterized Quantum Circuits (PQCs) \cite{19, 20}. Quantum Architecture Search (QAS) is a technique employed to automate the discovery of high-performance PQCs \cite{21, 22, 23, 24, 25, 26, 27, 28, 29, 30, add_1}. QAS aims to identify the optimal quantum circuit within a vast search space that encompasses all possible circuit configurations. Consequently, QAS algorithms encounter significant computational challenges due to the exponential growth of the search space, particularly when scaling to large quantum circuits \cite{31}. Therefore, developing strategies to efficiently reduce the search space is crucial for improving the efficiency and scalability of QAS.

Several methods have been proposed to reduce the search space for QAS algorithms. A common approach involves imposing strict constraints on the design of PQCs \cite{32}, such as limiting quantum gates to act on fixed qubits rather than arbitrary ones. However, these methods indiscriminately discard numerous potential PQCs without a targeted approach, which may inadvertently exclude optimal circuit configurations. Another approach focuses on identifying modular components of quantum circuits and constructing larger circuits by repeating these components \cite{27, 33}. This method effectively manages large search spaces but may results in circuits with increased depth. Additionally, a preprocessing technique has been proposed to filter out low-performance quantum circuits in QAS \cite{34}. While this approach significantly improves the search efficiency of QAS, it introduces additional computational complexity.

In this paper, we propose a novel framework to significantly reduce the search space in QAS. Our approach is inspired by Neural Architecture Search (NAS), which aims to automatically identify high-performance neural network architectures for a specific task \cite{36}. It has been established that the performance of Convolutional Neural Networks (CNNs) identified through NAS remains relatively robust to the random replacement of specific operations, such as convolutional layers \cite{37}. Building on this observation, we explore whether a similar phenomenon occurs in VQA, where the circuit topology plays a more critical role than specific gate types in determining performance. This motivates us to investigate a method for decoupling the QAS process into two independent components: topology and gate types. This decoupling strategy allows QAS to avoid exploring gate-type configurations associated with low-performance topologies.

Through extensive numerical simulations, we validate that topology indeed plays a more dominant role than gate types in the search for high-performance quantum circuits.
Therefore, we propose the Topology-Driven Quantum Architecture Search (TD-QAS) framework, which prioritizes an independent search for high-performance topologies, followed by an efficient fine-tuning of gate types. 
The TD-QAS framework effectively decouples the original search space into two distinct spaces —topology and gate types —thereby significantly reducing the overall search space.
Furthermore, we observe that during the gate-type optimization phase, QAS can inherit trainable parameters from the topology search phase. This inheritance method eliminates the need to train the gate-type optimization phase from scratch, accelerating convergence.
We performed numerical simulations on various tasks in both noiseless and noisy scenarios. The results demonstrate that the proposed TD-QAS framework successfully identifies high-performance quantum circuits while maintaining significantly lower computational complexity compared to original QAS algorithms, which search both topology and gate types.
This study enhances our understanding of VQAs as well as broadens the applicability of QAS to more complex tasks, paving the way for more efficient and scalable approaches to quantum circuit design.

\newpage

\section{Background and related work}
\subsection{Quantum architecture search}
QAS is a promising technique for automating the search for high-performing PQCs, enabling integration of specific target tasks or additional constraints during the search process \cite{30}. It explores the expansive search space of potential quantum circuits by employing a search strategy to generate candidate circuits, followed by an evaluation strategy to assess their performance. The feedback from these performance evaluations is used to optimize the trainable parameters of the search model.
However, QAS faces significant challenges related to high computational complexity, which has spurred the development of various strategies to mitigate these costs.

Efforts to reduce the computational complexity of QAS have focused on three key aspects: search strategy, evaluation strategy, and search space optimization. The most basic search strategy is random search strategy, which often requires an impractically numerous iterations to identify an effective quantum circuit, causing excessive computational cost. Therefore, advanced search models have been introduced to enhance search efficiency or reduce training costs. These models include evolutionary algorithms, reinforcement learning \cite{21}, differentiable algorithms \cite{24}, and Bayesian optimization methods \cite{27}, among others.

Although advanced search strategies enhance search efficiency, they require substantial feedback for parameter training, leading to the evaluation of numerous quantum circuits; thus, more efficient evaluation techniques have been developed. These methods aim to accelerate the evaluation process while maintaining evaluation accuracy. Notable approaches include shared parameter evaluation (one-shot techniques) \cite{23}, performance predictor-based evaluations \cite{25,28}, and training-free predictors \cite{22}.

The extensive search space in QAS presents challenges in efficiently learning from exploration and accurately evaluating quantum circuit performance. Therefore, various techniques have been proposed to reduce the search space. These include imposing strict constraints on quantum gate operations \cite{31}, searching for small-scale circuit components to construct larger circuits \cite{27, 32}, and employing search space pruning methods \cite{33}. By narrowing the search space, these techniques improves QAS efficiency and scalability, enabling it to focus on more promising regions.

In this work, we implement the TD-QAS framework based on two representative QAS algorithms (QuantumSupernet \cite{23} and DQAS \cite{24}), to validate the improvements introduced by our approach over traditional QAS methods. The following subsections provide concise overview of the two baseline algorithms, while Section \ref{TDGT-QAS} describes how TD-QAS can be seamlessly integrated with them without altering their core mechanisms.

\subsubsection{QuantumSupernet}
\label{QuantumSupernet}

QuantumSupernet, proposed by Du et al. \cite{23}, is a representative QAS algorithm designed to efficiently explore large quantum circuit spaces using a one-shot neural architecture search paradigm. The core concept of QuantumSupernet is to construct a parameter-sharing supernet that encodes numerous candidate circuits within a unified structure. The supernet consists of multiple layers, each containing a set of single-qubit and two-qubit gates from a predefined native gate set. This layered representation offers a flexible yet efficient search space for quantum circuit design. To explore this space, QuantumSupernet employs a sub-sampling mechanism to extract candidate sub-circuits. The method supports both uniform random sampling and genetic algorithm-based sampling strategies. The performance of each sub-circuit is estimated through a shared-parameter evaluation mechanism, eliminating the need to train every candidate circuit from scratch. The shared parameters are jointly optimized within the supernet, enabling efficient reuse and rapid evaluation of multiple sub-circuits.

The overall workflow of QuantumSupernet comprises four stages: (1) initialization phase, in which the supernet architecture and native gate set are defined; (2) shared-parameter training phase, where sub-circuits are iteratively sampled and evaluated using the shared parameters, which are updated based on performance feedback; (3) architecture search phase, during which multiple sub-circuits are generated using either random or genetic sampling strategies, which are evaluated using shared parameters; (4) evaluation phase, where the best-performing sub-circuit is retrained from scratch to obtain its actual performance. This streamlined process enables QuantumSupernet to significantly reduce the computational cost of QAS while maintaining competitive accuracy across a variety of quantum tasks.

\subsubsection{DQAS}
\label{DQAS}

Differentiable Quantum Architecture Search (DQAS), proposed by Zhang et al. \cite{24}, is a flexible framework that incorporates differentiability into quantum circuit architecture search, enabling gradient-based optimization of circuit structures. DQAS distinguishes itself from other QAS algorithms by formulating the search for circuit structures as a differentiable process, thereby significantly enhancing search efficiency.

DQAS models quantum circuit design as a bi-level optimization problem, jointly optimizing both the circuit structure and the associated gate parameters. The core concept is to relax the discrete search space of circuit architectures into a continuous domain using a parameterized probabilistic search strategy. The search space consists of a pool of candidate operations (quantum gates) and a circuit encoding mechanism that enables flexible assignment of these operations to the architecture.
During each training iteration, DQAS samples a batch of circuit structures from the probabilistic model and assess their performance using a shared-parameter evaluation strategy. Gradients for both structure parameters (which define the probabilities of selecting specific gates at each layer) and shared parameters (which define the variational parameters within each gate) are computed and updated via automatic differentiation. This approach allows efficient, end-to-end learning of both the circuit layout and gate parameters.
The final architecture is determined by selecting the most probable circuit structure based on the learned probabilistic model. This structure is retrained from scratch to obtain its true performance. The overall DQAS workflow consists of five stages: initializing the gate pool, sampling circuit architectures, updating structure and gate parameters via gradient descent, selecting the optimal architecture, and retraining it from scratch. Further details are available in the original work \cite{24}.

\subsection{Numerical simulation tasks}
\label{task}
VQAs are fundamental to quantum computing, offering a wide range of important applications. In QAS, three representative benchmark tasks commonly used to validate algorithm effectiveness include ground state energy estimation for molecular systems, solving the MaxCut problem, and performing classification.

\subsubsection{Variational quantum eigensolver}
Calculating the ground state and the lowest energy of Hamiltonian is a fundamental problem in physics research \cite{38, 39, 40}. In quantum systems, the Hamiltonian's dimensionality increases exponentially with system size, posing significant computational challenges.  
Therefore, researchers have employed the Variational Quantum Eigensolver (VQE), which offers exponential speedup, to address this issue \cite{41, 42, 43, 44}.
The key to solving this problem is to prepare the ground state $| \varphi_\text{0} \rangle$ of the Hamiltonian $H$ and compute the lowest energy $E_\text{0}$, as expressed in (\ref{eq:E_0}):

\begin{equation}
E_0 = \frac{\langle \varphi_0^* | H | \varphi_0 \rangle}{\langle \varphi_0^* | \varphi_0 \rangle},
\label{eq:E_0}
\end{equation}
As the $|\varphi_0 \rangle$ is difficult to prepare, VQE approximates it by constructing a states  $|\varphi^n \rangle$ by a PQC $U(\theta)$, as shown in (\ref{eq:state}):
\begin{equation}
|\varphi^{n}\rangle = U(\theta) | 0^n \rangle,
\label{eq:state}
\end{equation}
Here, $|0^n \rangle$ represents the all-zero quantum state, which serves as the initial state and processed by the $U(\theta)$.
The goal of VQE is to iteratively adjust the parameters $\theta$ such that $| \varphi^n \rangle$ gradually approaches $|\varphi_0 \rangle$. Therefore, calculating the ground state and the lowest energy can be transformed into finding a suitable $\theta$ that minimizes (\ref{eq:opt}):
\begin{equation}
\theta^* = \arg \min_{\theta} L(\theta) = \langle 0^n | U^{\dagger}(\theta) H U(\theta) | 0^n \rangle,
\label{eq:opt}
\end{equation}

The process of VQE can be summarized as first preparing the quantum state $|\varphi^n \rangle$  through $U(\theta)$, then measuring the energy $E$ of the state $|\varphi^n \rangle$, and finally updating $\theta$ until the minimum $E$ is obtained.

\subsubsection{Variational quantum algorithm for the MaxCut problem}
The MaxCut problem is a classic combinatorial optimization problem. Consider a graph $G = (V, E)$, the objective is to partition its nodes into two distinct sets to maximize the number of edges connecting these sets.
The objective function of MaxCut can be written as:
\begin{equation}
C(z) = \frac{1}{2} \sum_{(i, j) \in E} \left( 1 - z_i z_j \right),
\label{eq:MaxCut_objective}
\end{equation}
where $z_{i},z_{j} \in \{+1,-1\}$ and $z_i$ represents the subset to which vertex $i$ belongs. 
The Quantum Approximate Optimization Algorithm (QAOA) addresses the MaxCut problem by utilizing alternating optimization to approximate the optimal solution, offering the potential for quantum speedup \cite{1}. 
Consequently, the problem Hamiltonian $H_c$ that encodes the MaxCut objective can be expressed as:
\begin{equation} 
H_C = \frac{1}{2} \sum_{(i, j) \in E} \left( I - Z_i Z_j \right),
\label{eq：HC} 
\end{equation}
where $Z_i$ and $Z_j$ are Pauli-Z operators acting on qubits corresponding to the nodes $i$ and $j$ of edge $(i, j) \in G$. These operators are used to encode the connectivity of the graph in the quantum circuit.
The PQC of QAOA is driven by the problem. By adjusting PQC's parameters, QAOA can progressively approximate the optimal solution.
In this study, we employed a VQA searched by the QAS algorithm to replace ansatz of QAOA for solving the MaxCut problem. The solving process is similar to VQE, except for the Hamiltonian $H$.

\subsubsection{Quantum Neural Network for Quantum State Classification}
Quantum Machine Learning (QML) is an advanced field that integrates quantum computing with machine learning, aiming to exploit the unique properties of quantum mechanics to enhance data analysis efficiency. 
In this study, we employed a Quantum Neural Network (QNN) to address a binary quantum state classification problem. 
We utilized the quantum entanglement dataset provided in Ref. \cite{46}, which consists of quantum states characterized by different levels of Concurrence Entropy (CE). 
This dataset was utilized for training the QNN, which comprises an embedding layer, PQC, and a measurement layer. 
The embedding layer was implemented using angle encoding with $R_x$ rotation gates to map classical information to quantum states. The PQC component, searched by the QAS algorithm, processes the quantum states that contain the information from the dataset. Finally, the measurement layer extracts classical information from quantum states using Pauli-Z measurements, which are processed by a multilayer perceptron (MLP) to provide the binary classification result.

\section{Method}
\label{method}

\subsection{Motivation and overview}

\begin{figure}[!t]
\centering
\includegraphics[width=1\linewidth]{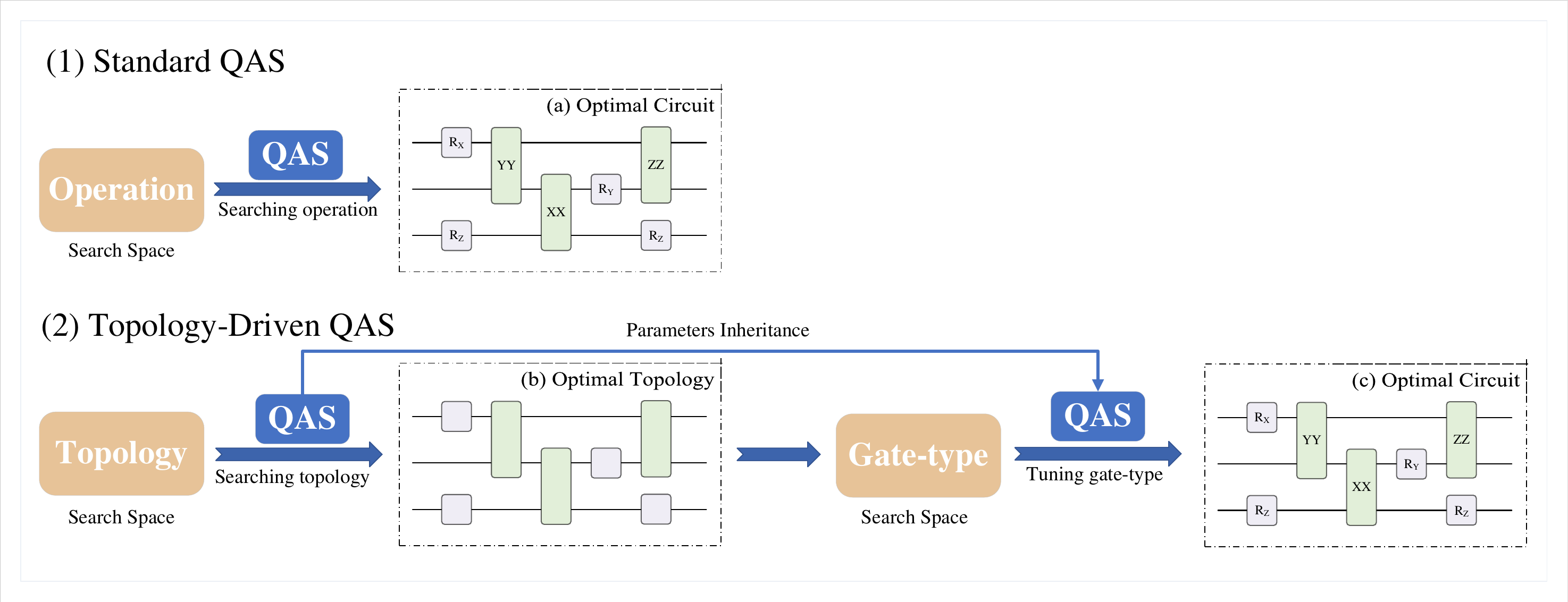}
\caption{
Comparison between standard QAS and Topology-Driven QAS (TD-QAS).  
(1) In standard QAS, the search must jointly explore both topology and gate types, causing a large and complex search space.  
(2) TD-QAS decouples the search into two sequential stages, which first identifies optimal topologies using symbolic placeholders, and then assigns gate types during a separate tuning phase.  
Here, (a) and (c) represent complete quantum circuits with both topology and gate-type information, while (b) depicts a topology-only representation using placeholders.
}

\label{fig1}
\end{figure}

Standard QAS algorithms explore a vast search space that encompasses both circuit topology information, such as qubit involvement, and gate-type information.
For instance, a fundamental element of a quantum circuit, such as $[R_x0]$, specifies that the $R_x$ gate acts on qubit 0.
Consequently, an entire circuit can be represented as a sequence of these elements, such as [$R_x0$, $R_z2$, $YY0$, $XX1$, $R_y1$, $ZZ0$, $R_z2$], as illustrated in Figure \ref{fig1} (a).
However, this approach significantly increases the search difficulty in QAS. Specifically, heuristic search strategies and evaluation mechanisms must simultaneously explore both topology and gate-type information. This joint exploration across an extremely large search space results in substantial computational complexity and inefficiency. NAS has demonstrated that high-performance CNNs remain unaffected by the random replacement of specific operations \cite{45}. If a similar phenomenon occurs in VQA, decoupling topology and gate-type searches could enhance QAS in discovering high-performance quantum circuits. This method can significantly reduce the search space by eliminating the need to search for gate configurations within low-performance topologies.

Extensive numerical simulations, as discussed in Section \ref{Numerical_Simulation1}, reveal that high-performance quantum circuits exhibit only slight performance changes when quantum gate types are randomly substituted, provided that the circuit topology remains unchanged. 
Remarkably, 76\% of the quantum circuits exhibited no significant change in performance after their gate types were altered in noise scenario.
This observation indicates that identifying an optimal circuit topology is the primary step in searching for a high-performance VQA, followed by gate-type selection. Therefore, we introduce the Topology-Driven Quantum Architecture Search (TD-QAS) framework, as illustrated in Figure \ref{fig1} (b), which significantly reduces the search space size.
Our framework first employs a topology-driven (TD) phase, which prioritizes the identification of high-performance topologies. Subsequently, the framework utilizes a gate tuning (GT) phase to adjust the suitable gate types to enhance circuit performance. To circumvent the substantial computational complexity associated with training the QAS of the tuning phase from scratch, the tuning phase inherits parameters from the TD-QAS phase, enabling rapid convergence with only a few additional training epochs.

\subsection{Implementation of TD-QAS framework}

Presentation and evaluation strategy collectively enable topology search within QAS algorithms. We emphasized topology representation and a dedicated search space. At the algorithmic description level, circuit topology is defined as a sequence of placeholders, each representing a potential position for a quantum gate. The type of placeholder depends on the number of qubits the gate acts on. As shown in Figure \ref{fig1}(b), a topology might be represented as [$single$0, $single$2, $double$0, $double$1, $single$1, $double$0, $single$2], where single placeholders correspond to single-qubit gates, and double placeholders correspond to two-qubit gates. This configuration explicitly defines the topology search space for a task with a fixed number of qubits and quantum gates.

To facilitate the search for topology, it is essential to establish an evaluation method specific to topology. The straightforward approach to evaluate a topology's performance is to exhaustively enumerate or randomly generate numerous quantum circuits, using their performance metrics as proxies for the topology's effectiveness. However, this method becomes computationally prohibitive, especially as the search space expands.
We introduce a resource-efficient topology evaluation strategy called the Topology Instantiation Evaluation Method. In this approach, a topology instantiation is created by replacing single- and double-qubit placeholders with predefined gates, such as $R_X$ for single-qubit and $XX$ (or $CNOT$) for two-qubit operations. The performance of a topology instantiation provides an approximation of the topology's true performance. 
Considering the relatively limited impact of gate types on circuit performance within QAS, this method provides a reasonable and efficient solution to evaluate topology without incurring substantial computational complexity. Section \ref{Numerical_Simulation2} verifies the accuracy of the topology instantiation evaluation method in reflecting the true performance of the topology. This validation ensures that our evaluation strategy reliably estimates the effectiveness of different topologies, thereby supporting the efficiency of the TD-QAS framework.

The proposed topology representation and evaluation strategy collectively enable topology search within QAS algorithms. This topology search readily integrates into mainstream QAS algorithms without requiring any modifications to their core search mechanisms. Specifically, when the native gate set is limited to a single-qubit gate and one type of two-qubit gate, mainstream QAS algorithms transition from searching for quantum circuits to focusing on topology. 
This design aligns well with our proposed Topology Instantiation Evaluation method. Consequently, at the level of practical implementation, all topology representations can be evaluated through their instantiations.
After completing topology search, the TD-QAS framework proceeds to the gate-type search phase. The size of the gate-type search space depends on both the selected topology and the native gate set. 
The core mechanisms, including the evaluation strategy and search procedure, remain unchanged and are omitted here for brevity. In summary, the TD-QAS framework can be implemented on existing QAS methods with minimal modifications.

\subsection{Reduction of search space}
\label{33}
In this subsection, we analyze how the TD-QAS framework effectively reduces the search space, offering an intuitive perspective of its impact. 
First, the topology search space is typically determined by the number of placeholder types and the properties of the quantum circuit, such as the number of qubits and gates.
The gate-type search space is determined by the searched topology and the native gate set, which is typically defined by manual design or hardware constraints. 
To illustrate the efficacy of the TD-QAS framework in reducing the search space, consider a three-qubit quantum circuit consisting of seven quantum gates ($N_{\text{qubit}} = 3$, $N_{\text{gate}} = 7$) with ring connectivity among all qubits. We assume the qubit connectivity is ring connectivity, and there are $N_{\text{qubit}}$ possible positions for placing either single-qubit or two-qubit gates. Suppose we set the native gate set \( A = \{R_x, R_y, R_z, XX, YY, ZZ\} \), thus \( |A[\text{single}]| = 3 \) and \( |A[\text{double}]| = 3 \), where \( |A[\text{single}]| \) denotes the number of single-qubit gates in the native gate set. 
In classical QAS, the number of possible operations per layer is 18 \( (N_{\text{qubit}} \cdot (|A[\text{single}]| + |A[\text{double}]|)) \), and the search space size is \( 18^7 \) \( \left[N_{\text{qubit}} \cdot (|A[\text{single}]| + |A[\text{double}]|)\right]^{N_{\text{gate}}} \). 
The TD-QAS framework employs two types of placeholders, with placeholders in each layer having 6 \( (2 \cdot N_{\text{qubit}}) \) possible combinations. Thus, the topology search space size becomes \( 6^7 \) \( (|(N_{\text{qubit}} \cdot 2)^{N_{\text{gate}}}|) \). 
The gate-type search space is \( 3^7 \) \( (|A[\text{single}]|^x \cdot |A[\text{double}]|^{7-x}) \), where \( x \) represents the number of single placeholders in the topology.
The TD-QAS framework decouples the operation search space into a topology search space and a gate-type search space, reducing it from \( 18^7 \) to \( 6^7 \) and \( 3^7 \). The corresponding search space reduction ratio is approximately \( 2 \times 10^3 \), and the calculation formula is provided in (\ref{eq:Reduction_Ratio}). In this study, we use the reduction ratio $R$ to evaluate the compression efficiency of our framework.
\begin{equation}
R = \frac{S_{\text{proposed}}}{S_{\text{traditional}}}.
\label{eq:Reduction_Ratio}
\end{equation}

\subsection{Quantum computational costs}
\label{Quantum_Computational_Costs}
In this work, we introduce the TD-QAS framework, which is characterized by exceptionally low computational complexity. In QAS, quantum computational complexity is the primary consideration. Consequently, we utilize quantum computational costs to evaluate the computational overhead of QAS algorithms \cite{35}. Quantum computational costs estimate the execution time for all quantum circuits generated by QAS algorithms on actual devices.
The execution time $T$ for an individual quantum circuit can be expressed as:
\begin{equation}
T = T_p + (circuit\ depth) \cdot T_G + T_M,
\label{eq:quantum_time}
\end{equation}
where $T_P$ is the time to prepare the initial state, $T_g$ is the average duration of a quantum gate, and $T_M$ is the time required to measure the qubits. 
Here, $T_p + T_M$ is 1 microsecond, and the average gate time $T_G$ is set to 0.01 microseconds.
Circuit depth represents the maximum length of the directed path from input to output in the quantum circuit, directly influencing the coherence time required to complete the algorithm. In our numerical simulations, we will compare the quantum computational costs between TD-QAS framework and original QAS algorithms, aiming to evaluate their respective efficiencies.

\section{Numerical simulation}
\label{Numerical_Simulation}
In this section, we conduct extensive numerical simulations to evaluate the proposed TD-QAS framework across three representative tasks both noiseless and noisy scenarios.
Specifically, three representative tasks include estimating ground state energies for molecular systems, solving the MaxCut problem on 100 Erdős-Rényi (ER) graphs with 10 nodes, and performing an 8-qubit quantum state classification task.  In the noisy scenario, we adopted depolarizing and bit-flip noise models, with parameters set to 0.01 for single-qubit depolarizing, 0.001 for two-qubit depolarizing, and 0.01 for bit-flip noise.

In the following sections, we first validated that high-performance quantum circuits experience only minor performance variations when quantum gate types are randomly replaced, provided the circuit topology remains unchanged. Thus, prioritizing the search for circuit topology before fine-tuning gate types is a reasonable approach.
Secondly, we validated that the topology instantiation evaluation method effectively assess circuit topology performance, offering an evaluation strategy that enables QAS algorithms to efficiently search for low-complexity circuit topologies.
Finally, we verify the superiority of our TD-QAS framework by integrating it with two representative QAS algorithms, which is QuantumSupernet \cite{23} and DQAS \cite{24}.

\subsection{Core hypothesis of TD-QAS: topology dominates the performance of circuit}
\label{Numerical_Simulation1}

This subsection aims to validate the core hypothesis of the TD-QAS framework, which posits that quantum circuit topology has a greater influence on VQA performance than quantum gate types.
This hypothesis is motivated by our preliminary observations, suggesting that for high-performing circuits, maintaining a fixed topology while randomly modifying the quantum gate-type has minimal impact on overall performance.
To verify the generality of this phenomenon, we conducted a series of numerical simulations. Specifically, we first describe the design and settings of numerical simulations. Thereafter, we validate the core hypothesis using a VQE task under both noiseless and noisy scenarios. Finally, we extend the evaluation to additional tasks, including the MaxCut problem and quantum state classification, to further demonstrate the robustness of the hypothesis across different task.

\begin{figure}[!t]
\centering
\includegraphics[width=1\linewidth]{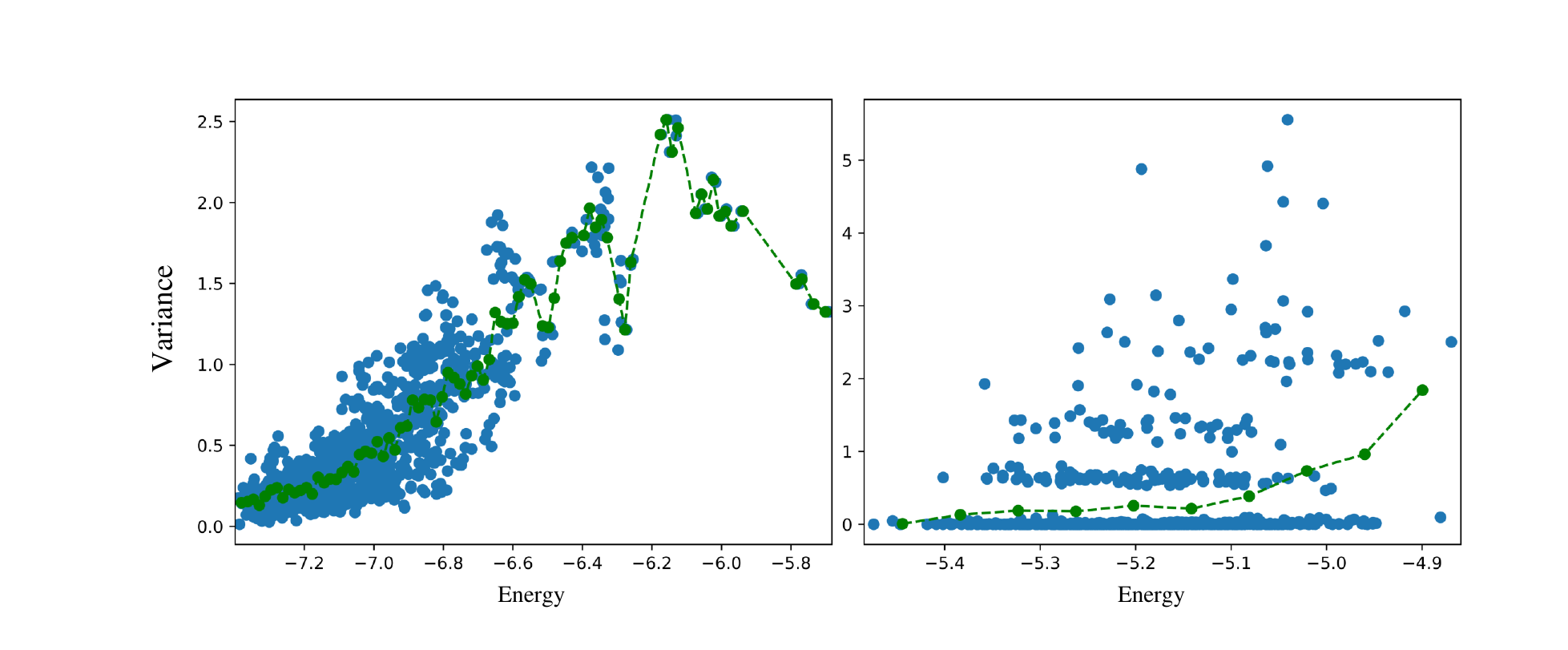}
\caption{Verification of the core hypothesis on the VQE task under noiseless (left) and noisy (right) conditions. The horizontal axis shows the original circuit’s performance, while the vertical axis indicates the performance difference between the original circuits and their gate-modified variants. Blue dots represent individual performance differences; the green line denotes the average difference across performance levels.}
\label{VQE1_noise}
\end{figure}

We outline the numerical simulation process, comparing the performance of the original quantum circuits with their gate-modified variants. Specifically, we first randomly generated 1,000 original quantum circuits. Each quantum circuit consisted of multiple quantum gates chosen from the native gate set $\{R_x, R_y, R_z, XX, YY, ZZ\}$, with qubit connectivity arranged in a ring connectivity.
For each original circuit, we generated 10 variants by randomly modifying 20\% of the gate types, ensuring that each replaced gate remained within the same class (i.e., single-qubit gates replaced by other single-qubit gates), thereby preserving the topology.
We selected the best performance from the five training runs as the true performance.
This performance dataset consist of pairs, each containing the original circuit performance and that of its corresponding variants [$p_i$, $v_i^j$], where $i$ represents the original circuit ($i \in [1, 1000]$) and $j$ represents the variant circuit ($j \in [1, 10]$).
To quantify the performance differences between each original circuit and its variants, we computed the mean squared error (MSE), yielding pairs of original performance and performance difference [$p_i$, $d_i$] as defined (8): 

\begin{equation}
d_i = \frac{1}{nm} \sum_{i=1}^{n} \sum_{j=1}^{m} \left( p_i - p_i^j \right)^2,
\label{eq:formula}
\end{equation}

We first evaluated this hypothesis on a VQE task (6-qubit TFIM) under both noiseless and noisy scenarios. Specifically, both the original circuits and their variants were configured to contain 35 quantum gates. The numerical simulation results are visualized in Figure~\ref{VQE1_noise}, where each blue dot represents a circuit-variant pair, and the green curve shows the average $d_i$ over performance intervals (100 and 10 bins, respectively). To better observe this trend under the noisy scenario, we excluded extreme outliers with performance below -5. 
In both noiseless and noisy scenarios, we observe a consistent trend: as circuit performance improves, the impact of gate-type changes diminishes. Remarkably, in the noisy VQE task, 76\% of circuits exhibited minimal performance change (MSE $<$ 0.1) after gate types were altered. We attribute this phenomenon to the influence of noise, which overwhelms the relatively small effects caused by changes in gate types. These findings provide strong support for the core hypothesis.

\begin{figure}[!t]
\centering
\includegraphics[width=1\linewidth]{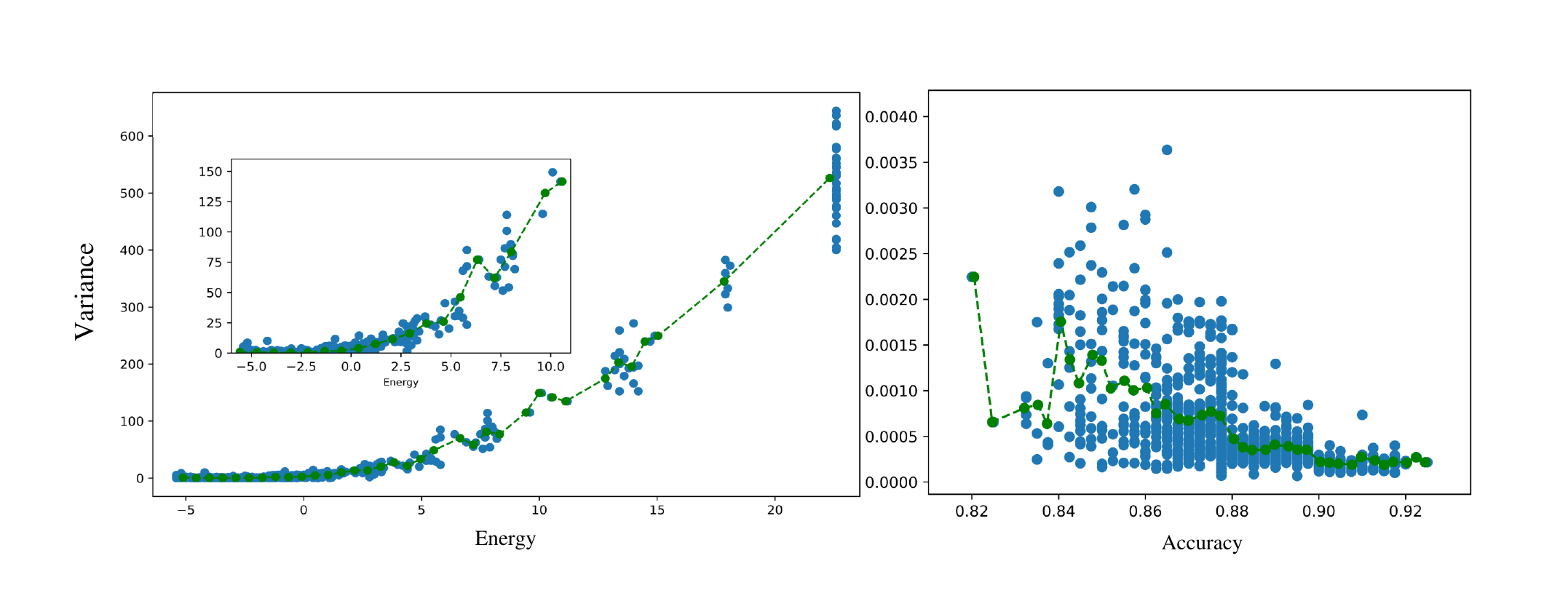}
\caption{Verification of the core hypothesis on solving the MaxCut problem (left) and quantum state classification task (right). The horizontal axis represents the original circuit’s performance, and the vertical axis represents the performance difference between the original circuits and their gate-modified variants. Blue dots represent individual performance differences, while the green line shows the average difference across different performance levels.}
\label{maxcut-classification1}
\end{figure}

To further assess the robustness of the hypothesis, we extended the simulation protocol to two additional tasks: solving the MaxCut problem and performing quantum state classification. For each task, the original circuits and their variants contained 35 or 50 quantum gates, and their performance was evaluated under noiseless scenarios. The results, shown in Figure~\ref{maxcut-classification1}, demonstrate a similar trend to that observed in the VQE experiments, reinforcing the generality of our core hypothesis across different tasks.

\subsection{Topology instantiation evaluation method}
\label{Numerical_Simulation2}

Existing QAS methods lack a dedicated approach for evaluating topology. The ideal method to evaluate a topology's performance is to exhaustively analyze all potential quantum circuits or numerous circuits, and use their average performance as a proxy for topology effectiveness. However, this approach is impractical owing to its high computational complexity.
As observed in Section \ref{Numerical_Simulation1}, we found that random gate substitutions have minimal impact on performance for high-performance circuits. Therefore, we propose a highly resource-efficient topology evaluation method called the Topology Instantiation Evaluation. This method represents the topology using topology instantiation, where the corresponding placeholders are replaced with specific quantum gates.
To validate the effectiveness of this method, we computed the correlation between the approximate true performance and the performance of the topology instantiation.

We first generated 100 circuit topologies. For each topology, we created 100 quantum circuits by randomly selecting quantum gates from the set \( \{R_x, R_y, R_z, XX, YY, ZZ\} \). For each topology, we initialized parameters for 100 corresponding circuits, trained them, and use their average performance as the approximate true performance, denoted as \( \bar{y} \). 
Subsequently, for each of topologies, we replaced the single-qubit and two-qubit placeholders with \( R_x \) gates and \( XX \) gates, respectively. This process generated one instantiation circuit for each topology. Similarly, we calculated the performance of these instantiations, denoted as \( y^{\prime} \). Finally, we computed the Pearson correlation coefficient between the true performance, \( y \), and the instantiation performance, \( y^{\prime} \).

\begin{figure}[!t]
    \centering
    \includegraphics[width=0.7\linewidth]{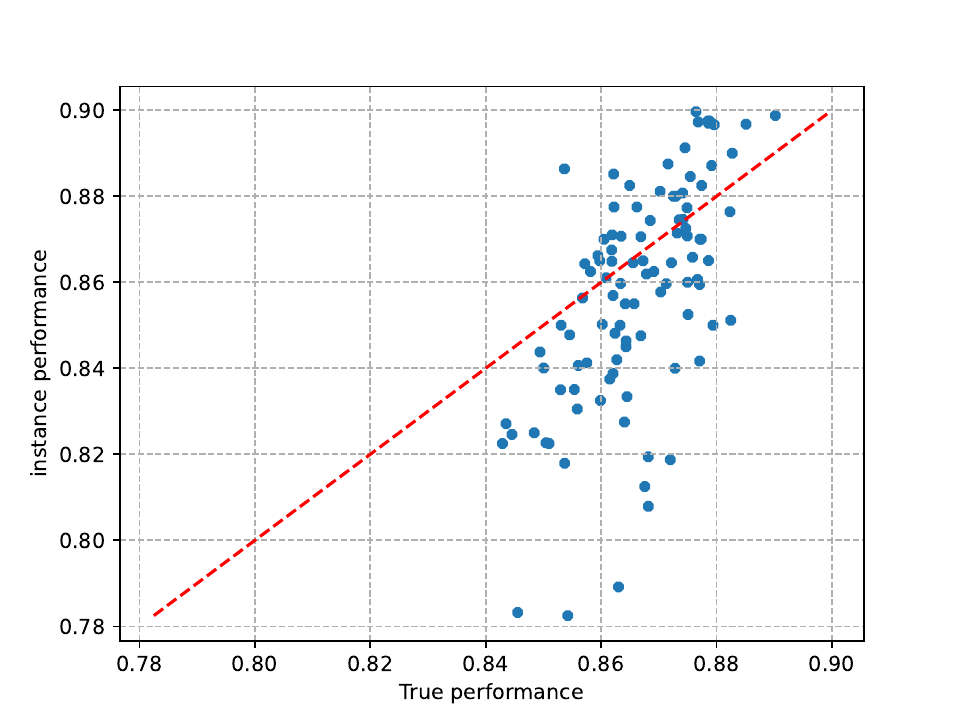}
    \caption{
    Correlation between true performance and instantiation-based approximation in the state classification task. Each point represents a topology, where the x-axis indicates the true performance (averaged over 100 gate-randomized circuits), and the y-axis shows the performance estimated by the Topology Instantiation Evaluation Method.
    }
    \label{fig:correlation}
\end{figure}

The numerical simulations yielded correlation values of 0.68, 0.59, and 0.62 for the three tasks. We present the results for one task in Figure \ref{fig:correlation}. For comparison, the widely used performance predictor achieved a correlation of 0.71 \cite{25}, although it required considerable complexity to prepare the training datasets. Thus, our method provides a reasonably accurate assessment of topology's performance while maintaining significantly lower complexity.

To further evaluate the robustness of the proposed Topology Instantiation Evaluation method, we conducted an additional experiment on the VQE task. In particular, we evaluated two alternative combinations of single- and double-qubit gates (\( R_y \) with \( YY \) and \( R_y \) with \( XX \)), measuring the instantiation performance and correlating it with the true average performance. The correlation values obtained with these gate sets were 0.68 and 0.65, respectively, which are closely aligned with the correlation of 0.68 observed when using the \( R_x \) and \( XX \) gates. These results indicate that the accuracy of the proposed evaluation method remains unaffected by the specific gate types used for instantiation.

\subsection{TD-QAS framework}
\label{TDGT-QAS}
In the previous sections, we discussed the challenges faced by existing QAS algorithms, including the vast search spaces and excessive computational costs. Therefore, we propose the TD-QAS framework, which aims to enhance the efficiency of QAS by decomposing the operational search space into separate topology and gate-type search spaces.
To assess the effectiveness of the TD-QAS framework, we compare several metrics from both the TD and GT phases with those of standard QAS algorithms. These metrics include the size of the search space, performance in specific tasks, circuit properties, and overall quantum computational costs.
In the following subsections, we apply the TD-QAS framework to two representative QAS algorithms and analyze their performance. 

\subsubsection{TD-QuantumSupernet framework}
\label{TDGT-QuantumSupernet}
In this subsection, we detail the implementation of the TD-QAS framework using QuantumSupernet, followed by a series of numerical simulations.
First, we explain how to implement the TD and GT phases using QuantumSupernet. The core algorithm of QuantumSupernet is described in Section \ref{QuantumSupernet}. Subsequently, we define the hyperparameters for various numerical simulation tasks, highlighting their impact on the comparison variables. Finally, we set the hyperparameters, execute the algorithm, and analyze the results of the numerical simulations. For the simulations, we select the same VQE and classification tasks used in QuantumSupernet.

\paragraph{Implementation of TD-QuantumSupernet}
Here, we describe the implementation of the TD-QAS framework within the QuantumSupernet algorithm. 
The TD phase begins by defining the topology search space based on the supernet structure and the native gate set. 
The number of qubits ($N_{\text{qubit}}$) and the number of layers ($N_{\text{layer}}$) are specified by the task.
Each layer supports up to $N_{\text{qubit}}$ quantum gates.
The native gate set is defined as $\{R_x, I, CNOT, CI\}$, where $CI$ refers to the Controlled-Identity Gate. If a sampled topology includes an $I$ or $CI$ gate at a specific location, no quantum gate will be placed there, and the GT phase will bypass fine-tuning for that position.
Subsequently, we utilize random search strategy to explore potential high-performance topologies. 
Each quantum circuit is evaluated using shared parameters, with feedback guiding the updates to these parameters. 
Specifically, $N_{\text{experts}}$ shared parameters are used to assess the performance of each topology, allowing for more precise evaluations. 
During the initial $T_{\text{warm}}$ iterations, a randomly selected expert from the $N_{\text{experts}}$ pool evaluates the topology instantiation and updates the shared parameters accordingly. 
Following the warm-up phase, all $N_{\text{experts}}$ contribute to the evaluation, but only the parameters with the best performance are updated.
When the shared parameters have converged, we evaluate $N_{\text{search}}$ topology instantiations, selecting the optimal topology.
In the gate-type (GT) phase, the search space is determined by the selected topology and the gate set $A_\text{g}$. To expedite training, the shared parameters from the TD phase are directly inherited and adjusted to accommodate the new gate configuration. Specifically, this inheritance mechanism involves replicating the shared parameters from the TD phase based on the number of single-qubit and two-qubit gates available in the native gate pool. To enhance the accuracy of the evaluation, we performed additional training on the inherited parameters for a few extra iterations, denoted as $T_{\text{extra}}$, which further refines the assessment of specific circuit performances.

Based on the defined hyperparameters, we compute key comparison variables, including the size of the search space and the total quantum computational cost(QCC).
As the supernet consistently applies both single-qubit and two-qubit gates  across all qubits at each layer, the original search space is calculated as $(|A[\text{single}]|^{N_{\text{qubit}}}\cdot |A[\text{double}]|^{N_{\text{qubit}}} )^{N_{\text{layer}}}$, where $|A[\text{single}]|$ denotes the number of single-qubit gates in the native gate set. 
The topology search space is defined as $(2^{N_{\text{qubit}}})^{N_{\text{layer}}}$, where $2^{N_{\text{qubit}}}$ accounts for all possible combinations of single- and two-qubit placeholders per layer.
The gate-type search space, conditioned on a specific topology, is expressed as $|A[\text{single}]|^{N_{\text{layer}}-\text{x}}\cdot|A[\text{double}]|^{\text{x}}$,
where $x$ denotes the number of double-qubit placeholders in the selected topology.
The quantum computational cost (QCC) for QuantumSupernet and the TD phase is given by $\sum_{\text{1}}^{T_{\text{warm}}} T_\text{i} + \sum_{T_{\text{warm}+1}}^{T} T_\text{i}*N_{\text{experts}} + \sum_{\text{1}}^{N_{\text{search}}} T_\text{i}*N_{\text{experts}}$
The QCC for the GT phase is denoted by $\sum_{\text{1}}^{N_{\text{search}}} T_\text{i} \cdot (1+T_{\text{extra}})$.
Here, $T_i$ represents the quantum computational cost of a single circuit evaluation, as defined in Equation~\ref{eq:quantum_time}.

\paragraph{Numerical simulation on the VQE task}
In this section, we evaluate the performance of the TD-QAS framework through numerical simulations. First, we replicate the ground state energy estimation task for a 4-qubit hydrogen molecule, which is consistent with the simulation in Ref. \cite{23}. While this small-scale task provides valuable insight, it does not fully demonstrate the potential of the TD-QAS framework. Therefore, we extend the validation to larger tasks, including the 5-qubit Heisenberg model.

For the simulation task estimating the ground state energy of hydrogen (which has a ground state energy of -1.13618 Hartree), the Hamiltonian construction is detailed in QuantumSupernet \cite{23}. Before performing the numerical simulations, we define the hyperparameters and compare the search space sizes of the QuantumSupernet and TD-QAS methods. Specifically, we set the number of qubits ($N_{\text{qubit}}$) to 4 and the number of layers ($N_{\text{layer}}$) to 3.
For the QuantumSupernet, the native gate set $A$ is defined as $\{I, R_y, R_z, CNOT, CI\}$. In the TD-QAS framework, the gate sets are defined as $A_\text{t} = \{R_y, I, CNOT, CI\}$ for the topology-driven phase and $A_\text{g} = \{R_y, R_z, CNOT\}$ for the gate tuning phase.
Under these settings, the search space for the QuantumSupernet is calculated as $(3^4 \cdot 2^4)^3$. The search space for the TD phase is calculated as $(2^4 \cdot 2^4)^3$, while the GT phase operates within a search space ranging from 1 to $2^{12}$. This decoupling yields a search space reduction ratio of $2.5 \times 10^1$ compared to the original QuantumSupernet.
Owing to the reduced search space, we employ smaller values for the number of experts ($N_{\text{experts}}$) and the total number of iterations ($T$) to enhance efficiency. Specifically, for the QuantumSupernet, we set $N_{\text{experts}} = 5$, $T = 500$, $T_{\text{warm}} = 200$, and $N_{\text{search}} = 500$. In the TD phase, these parameters are adjusted to $N_{\text{experts}} = 1$, $T = 200$, $T_{\text{warm}} = 100$, and $N_{\text{search}} = 500$. In the GT phase, we set $T_{\text{extra}} = 1$. Consider the small problem size and the limited size of the native gate set, an exhaustive search is feasible in the GT phase, thus we omit $N_{\text{search}}$ for this stage.

A comparison of the numerical simulations for these methods is presented in Table \ref{tab:simulation_results_hydrogen}. In the noiseless scenario, the quantum circuit obtained in the TD and GT phases performs comparably to the QuantumSupernet. However, the circuit depth is reduced to 74.1$\%$, the number of parameterized gates to 32.6$\%$, and the total number of gates to 45.3$\%$ of the QuantumSupernet. Consequently, the quantum computational cost is reduced to approximately 19$\%$ of the QuantumSupernet. This demonstrates the effectiveness of independently searching for the topology and gate types, as the topology search alone successfully complete the task, achieving chemical accuracy with an estimation error below 0.0016 Hartree.
In the noisy scenario, the TD-QAS framework identifies higher-performance circuits, likely owing to the reduced circuit depth. Overall, the results from both noiseless and noisy scenarios confirm that prioritizing the search for circuit topology, followed by fine-tuning the gate types, enables the TD-QAS framework to find high-performance circuits with lower quantum computational costs.

\begin{table}[!t]
\footnotesize
\caption{Simulation Results for the Ground State Energy Estimation of Hydrogen. The "Property" column shows the layer depth, number of parameterized gates, and total gate count of the searched circuits. "QuantumSupernet-5" and "TD-1" refer to the method with $N_{\text{experts}}$ as indicated.}
\label{tab:simulation_results_hydrogen}
\tabcolsep 25pt 
\begin{tabular*}{\textwidth}{lcccc}
    \toprule
    Method & Energy & Property & QCC & Scenario \\
    \hline
    QuantumSupernet-5 & -1.13610 & 6.2, 9.2, 16.1 & 4471.4 & noiseless \\
    TD-1 & -1.13572 & 4.6, 3.0, 7.3 & 846.2 & noiseless \\
    GT-1 & -1.13572 & 4.6, 3.0, 7.3 & 16.7 & noiseless \\
        \hdashline  
    QuantumSupernet-5 & -0.97079 & 3.8, 7.4, 12.8 & 4327.4 & noise \\
    TDGT-1 & -1.09169 & 4.0, 3.0, 6.4 & 841.6 & noise \\
\bottomrule
\end{tabular*}
\end{table}

To further demonstrate the capabilities of the TD-QAS framework for larger-scale problems, we applied it to the ground state energy estimation of the 5-qubit Heisenberg model. The Hamiltonian for the Heisenberg model is defined as $\sum_{\text{i=1}}^{\text{n}} \left( X_\text{i} X_{\text{i+1}} + Y_\text{i} Y_{\text{i+1}} + Z_\text{i} Z_{\text{i+1}} \right) + \sum_{\text{i=1}}^{\text{n}} Z_\text{i}$ 
, and the ground state energy is -8.47213.
The hyperparameters for this simulation are defined as
$N_{\text{qubit}}$=5, $N_{\text{layer}}$= 6, $A$=\{$I$, $R_x$, $R_y$, $R_z$, $CNOT$, $CI$\}, $A_\text{t}$=\{$R_y$, $I$, $CNOT$, $CI$\}, and $A_\text{g}$=\{$R_x$, $R_y$, $R_z$, $CNOT$\}.
With these settings, the search space for the QuantumSupernet is calculated as $(4^5 \cdot 2^5)^6 = 2^{80}$, while the search space for the TD and GT phases is $2^{60}$ and $1 \sim 3^{30}$, respectively. This results in a search space reduction ratio of approximately $1 \times 10^7$.
To evaluate the performance of the TD-QAS framework and compare it to the QuantumSupernet, we executed simulations with different values of $N_{\text{experts}}$. Specifically, we set $N_{\text{experts}} = 1$ and $N_{\text{experts}} = 5$. The hyperparameters for the QuantumSupernet are $T = 500$, $T_{\text{warm}} = 200$, and $N_{\text{search}} = 500$. For the TD-QAS framework, the parameters are adjusted as $T = 200$, $T_{\text{warm}} = 100$, and $N_{\text{search}} = 300$. In the GT phase, we set $T_{\text{extra}} = 1$ and $N_{\text{search}} = 100$.

\begin{table}[!t]
\footnotesize
\caption{Simulation Results for the Ground State Energy Estimation of Heisenberg. The property presents layer depth, number of parameterized Gates, and total gate number of searched circuits. QuantumSupernet-5 and TD-1 present method-$N_{\text{experts}}$}
\label{tab:simulation_results_heisenberg}
\tabcolsep 25pt 
\begin{tabular*}{\textwidth}{lcccc}
    \toprule
        Method & Energy & Property & QCC & Scenario \\
        \hline
        QuantumSupernet-1 & -7.64036 & 16.0,38.2,51.6 & 1155.1 & noiseless \\
        TD-1 & -8.00899 & 14.8,18.4,32.6 & 624.4 & noiseless \\
        GT-1 & -8.11899 & 14.8,18.4,32.6 & 332.6 & noiseless \\
        \hdashline 
        QuantumSupernet-5 & -8.22164 & 16.6,38.4,52.8 & 4873.2 & noiseless \\
        TD-5 & -8.01173 & 15.4,21.0,37.4 & 3945.2 & noiseless \\
        GT-5 & -8.14162 & 15.4,21.0,37.4 & 2126.9 & noiseless \\
        \hdashline  
        QuantumSupernet-5 & -7.82812 & 15.1,36.7,50.2 & 4893.2 & noise \\
        TDGT-1 & -8.02160 & 12.5,20.5,34.0 & 997.5 & noise \\
\bottomrule
\end{tabular*}
\end{table}

The results of the numerical simulations are presented in Table \ref{tab:simulation_results_heisenberg}.  With the same $N_{\text{experts}}$, the TD and GT methods outperform the QuantumSupernet. Therefore, the TD-QAS framework enhances the performance of the QuantumSupernet.
When comparing simulations with different numbers of experts ($N_{\text{experts}} = 1$ and $N_{\text{experts}} = 5$), we observe that the QuantumSupernet shows a higher dependency on $N_{\text{experts}}$. This is owing to various parameterized quantum gates in the circuits searched by QuantumSupernet, which results in greater sensitivity to the number of experts. However, the circuits found by the TD and GT methods have fewer parameters, resulting in more stable performance across varying values of $N_{\text{experts}}$. This highlights the TD-QAS framework's ability to reduce the training burden, which in turn optimizes the QCC.
Additionally, the circuits discovered in both the TD and GT phases exist within the operation search space, but were not identified by the QuantumSupernet. Therefore, the TD-QAS framework may help mitigate the risk of the QAS algorithm becoming trapped in local optima, enhancing the discovery of more optimal quantum circuits.

\paragraph{QNN for state classification task}
To assess the performance improvements of the TD-QAS framework on more complex tasks, we conducted a numerical evaluation using an 8-qubit quantum state classification task. Specifically, we employed a QNN to address a binary classification problem, where the input comprised quantum states with varying levels of entanglement, characterized by Concurrence Entropy (CE = 0.15 and CE = 0.45). The quantum dataset used in this experiment is based on the dataset from Ref. \cite{46}.

The hyperparameter settings for the numerical simulation are detailed as follows, 
$N_{\text{qubit}}$=8, $N_{\text{layer}}$= 6, $A$=\{$I$, $R_x$, $R_y$, $R_z$, $CNOT$, $CI$\}, $A_\text{t}$=\{$R_y$, $I$, $CNOT$, $CI$\}, and $A_\text{g}$=\{$R_x$, $R_y$, $R_z$, $CNOT$\}.
With these parameters, we calculate that the search space for the QuantumSupernet is $2^{144}$, while the search space for the TD and GT phases is $2^{96}$ and $1 \sim 3^{30}$, respectively. This results in a search space reduction ratio of approximately $2.6 \times 10^6$.
The simulation hyperparameters for the QuantumSupernet are set as $T = 500$, $T_{\text{warm}} = 200$, and $N_{\text{search}} = 500$. For the TD-QAS framework, we set $T = 200$, $T_{\text{warm}} = 100$, and $N_{\text{search}} = 300$ in the TD phase, while in the GT phase, $T_{\text{extra}} = 1$ and $N_{\text{search}} = 100$.

Table \ref{tab:numerical_simulation_results_calss} summarizes the simulation results for the state classification task. The TD-QAS framework successfully identifies higher-performance circuits compared to the QuantumSupernet, all while incurring lower quantum computational costs. 
Notably, this task presents a larger search space than any previously examined, and the TD-QAS framework significantly enhance QAS performance compared to earlier Numerical simulations. These results further emphasize the importance of the TD-QAS framework in effectively reducing the search space and enhancing search optimization.

\begin{table}[!t]
\footnotesize
\caption{Numerical Simulation Results for State Classification Task. The property presents layer depth, number of parameterized Gates, and total gate number of searched circuits. QuantumSupernet-5 and TD-1 present method-$N_{\text{experts}}$}
\label{tab:numerical_simulation_results_calss}
\tabcolsep 23pt 
\begin{tabular*}{\textwidth}{lcccc}
    \toprule
        Method & Success Rate & Property & QCC & Scenario \\
        \hline
        QuantumSupernet-5 & 0.83917 & 14.3 , 44.3 , 62.8 & 4800.6 & noiseless \\
        TD-1   & 0.84250 & 15.0 , 28.7 , 49.7 & 620.1  & noiseless \\
        GT-1   & 0.85017 & 15.0 , 28.7 , 49.7 & 224.9 & noiseless \\
        \hdashline  
        QuantumSupernet-5 & 0.80064 & 14.1 , 42.4 , 58.7 & 4792.2 & noise \\
        TDGT-1 & 0.81554 & 14.8 , 24.1 , 42.3 & 927.4 & noise \\
\bottomrule
\end{tabular*}
\end{table}

\subsubsection{TD-DQAS framework}
\label{TDGT-DQAS}
To validate TD-QAS framework’s generality and effectiveness across different QAS algorithms, we implement it using the DQAS algorithm and systematically analyze its performance. The core algorithm of DQAS has been detailed in Section \ref{DQAS}. Similar to Section \ref{TDGT-QuantumSupernet}, we first describe the implementation details of the TD-QAS framework using DQAS, followed by an introduction to the hyperparameter settings. Finally, we execute the QAS algorithms and analyze their results.

\paragraph{Implementation of the TD-DQAS}
Here, we outline the process of implementing the TD-QAS framework with DQAS.
DQAS aims to identify a high-performance quantum circuit that contains $N_{\text{gate}}$ quantum gates, where arbitrary quantum gates can be placed at any position.
In the TD phase, the native gate set is defined as $A_\text{t}$ is $\{R_x, XX \}$.
DQAS employs a search strategy base on probabilistic model to select a batch of $N_{\text{batch}}$ quantum circuits, each of which is composed of $N_{\text{gate}}$ gates chosen from the set $A_\text{t}$ and their corresponding qubit positions.
The performance of the selected circuits is evaluated using shared parameters, which are updated by feedback from the quantum circuit evaluation. This process continues until $N_{\text{train}}$ iterations or when it converge.
Subsequently, the probability model selects the optimal topology with the highest probability.
DQAS employs mini-batch gradient descent to optimize the model. Generally, models with greater complexity require a larger batch size.
In the GT phase, the shared parameters from the TD phase are inherited, and the probability model is trained from scratch. The model fine-tunes the quantum gates from the native gate set based on the searched topology. As the probability model only should consider gate types in the GT phase, the number of trainable parameters is small, allowing for a smaller $N_{\text{train}}$.

We calculate and compare several key metrics, including the search space and total quantum computational costs.
The search space of DQAS is defined by $(|A| \cdot N_{\text{qubit}})^{N_{\text{gate}}}$,
whereas the search space of the TD and GT phases is $(2 \cdot N_{\text{qubit}})^{N_{\text{gate}}}$ and $|A[\text{single}]|^{N_{\text{gate}} - x} \cdot |A[\text{double}]|^x$, respectively.
The QCC for all methods is computed as $\sum_{\text{i=1}}^{ N_{\text{train}}} \sum_{\text{j=1}}^{N_{\text{batch}}} T_\text{i}^\text{j}$.
In this section, we do not compare the properties of the quantum circuits, such as depth and the number of parameterized gates, as the gate placement rules in DQAS allow arbitrary single-qubit gates to act on any position.

\paragraph{Numerical simulation of TD-DQAS}

\begin{table}[!t]
\footnotesize
\caption{Numerical simulation results for TD-DQAS on three tasks. The "Performance" column shows the output metric for each task (energy, approximation ratio, or accuracy). QCC denotes quantum computational cost.}
\label{tab:performance_variables_DQAS}
\tabcolsep 40pt 
\begin{tabular*}{\textwidth}{lccc}
    \toprule
        Method & Performance & QCC & Scenario \\
        \hline
        \multicolumn{4}{ c }{VQE for TFIM} \\
        \hline
        DQAS & -6.38688 & 9972.5 & noiseless \\
        TD   & -6.53971 & 2391.4 & noiseless \\
        GT   & -7.06726 & 118.7  & noiseless \\
        \hdashline  

        DQAS & -6.12504 & 10786.5 & noise \\
        TDGT & -6.71821 & 3604.2 & noise \\
        \hline
        \multicolumn{4}{ c }{VQA for MaxCut} \\
        \hline
        DQAS & 0.77 & 11917.0 & noiseless \\
        TD   & 0.90 & 2322.8  & noiseless \\
        GT   & 0.93 & 361.8   & noiseless \\
        \hdashline  

        DQAS & 0.76 & 13835.9 & noise \\
        TDGT & 0.89 & 2945.6  & noise \\
        \hline
        \multicolumn{4}{ c }{QNN for State Classification} \\
        \hline
        DQAS & 0.78095 & 10603.4 & noiseless \\
        TD   & 0.78222 & 1750.7  & noiseless \\
        GT   & 0.85897 & 280.1   & noiseless \\
        \hdashline 

        DQAS & 0.77460 & 11894.4 & noise \\
        TDGT & 0.79811 & 2166.5  & noise \\
\bottomrule
\end{tabular*}
\end{table}

This section validates the capacity of the TD-DQAS framework across three tasks: the VQE task, the MaxCut problem, and the state classification task.
As numerical simulation setup and conclusions largely overlap with those discussed in earlier sections, we focus on presenting a comparative analysis of the results rather than repeating the full details.

For each task, we set the following hyperparameters,
 $A=A_\text{g}=\{R_x, R_y, R_z, XX, YY, ZZ\}$
$A_\text{t}=\{R_x, XX\}$. $N_{\text{train}}=500$, the $N_{\text{batch}}$ of DQAS, TD and GT phase is 32, 8, 8, respectively.
For the VQE task, we selected the 6-qubit TFIM, using the ground state energy as the performance metric. 
We set $ N_{\text{qubit}}=6$ and $N_{\text{gate}}=36$.
Therefore, the search space of DQAS, TD and GT phase is $36^{36}$, $12^{36}$ and $3^{36}$.
The search space reduction ratio is nearly $1.5 \times 10^{17}$.
In the MaxCut problem, we employed the ER model to randomly generate 100 distinct graphs, each containing 10 nodes, with an edge creation probability set to 0.5. 
The approximation ratio serves as the performance metric. 
We set $N_{\text{qubit}}=10$ , $N_{\text{gate}}=20$.
Therefore, the search space of DQAS, TD and GTphase is $60^{20}$,  $20^{20}$ and $3^{20}$.
The search space reduction ratio is nearly $3.5 \times 10^{10}$.
For the quantum state classification task, classification accuracy serving as the performance metric. 
We set $ N_{\text{qubit}}=8$ , $N_{\text{gate}}=30$.
Therefore, the search space of DQAS, TD, and GT phase is $48^{30}$, $16^{30}$ and $3^{30}$.
The search space reduction ratio is nearly $2 \times 10^{15}$.

Table \ref{tab:performance_variables_DQAS} presents detailed numerical results for each task, demonstrating that the TD-DQAS framework outperforms the DQAS algorithm across all tasks. 
Specifically, in the VQE, MaxCut, and quantum state classification tasks, the TD-QAS framework achieves higher performance with significantly reduced quantum computational costs. 
Therefore, decoupling the search process into two phases—topology search and gate-type fine-tuning—improves the efficiency of QAS algorithms, enabling them to obtain higher-performing quantum circuits while minimizing resource requirements.

\section{Conclusion and future work}
In this study, we proposed the TD-QAS framework, which significantly enhances the efficiency and scalability of QAS by decoupling the search process into two key steps: topology searching and gate-type fine-tuning. 
This novel approach minimizes the size and complexity of the search space, allowing for a more resource-efficient exploration. 
This method represents a novel paradigm for QAS, which can easily be integrated with both existing and future QAS algorithms without altering their core mechanisms.
Our experiments across various tasks, conducted in both noisy and noiseless scenarios, demonstrate that TD-QAS effectively discovers high-performing PQCs while reducing quantum computational costs. The TD-QAS framework was successfully implemented with two widely used QAS algorithms, i.e., QuantumSupernet and DQAS, demonstrating its compatibility. The numerical results indicate that TD-QAS improves QAS capacity while minimizing the risk of convergence to suboptimal circuits. To enhance reproducibility, the code for TD-QAS and all experiments will be made publicly available upon acceptance at: https://github.com/Sujun124/TD-QAS.

In future work, we first aim to extend TD-QAS to other QAS algorithms, thereby enhancing its utility and adaptability. Next, we will focus on optimizing the evaluation strategy, such as implementing a predictor-based evaluation method. 
Additionally, we plan to explore the scalability of TD-QAS for more complex tasks and scenarios. While our current study validates the core assumption and effectiveness of TD-QAS on three representative VQA tasks under both noiseless and noisy scenarios, this validation is limited to quantum machine learning tasks. Within the tasks explored, gate-type selection did not appear to play an equally or more critical role than topology. However, this assumption may not hold across all quantum computing settings. Investigating a broader range of tasks and environments remains an interesting direction for future research.
By refining these aspects, we aim to establish TD-QAS as a foundational approach for efficient and scalable QAS across various quantum computing applications.

\Acknowledgements{This work is supported by the National Natural Science Foundation of China (Grant Nos. 62372048, 62371069, 62272056)}

\end{document}